\documentclass[prd,a4paper,nofootinbib,
showpacs, twocolumn,
]{revtex4}
\usepackage{graphicx}
\usepackage{amsfonts}
\usepackage{amssymb}
\def\eq#1{{eq.~(\ref{#1})}}
\def\eqs#1#2{{eqs.~(\ref{#1})--(\ref{#2})}}

\def\vev#1{\left\langle #1\right\rangle}

\def\etal{{\it et al.}}

\def\hbar{\hspace{0pt}\raisebox{1pt}{$-$} \hspace{-7pt} h}

\def\5{\overline 5}

\newcommand{\be}{\begin{equation}}
\newcommand{\ee}{\end{equation}}
\newcommand{\bea}{\begin{eqnarray}}
\newcommand{\eea}{\end{eqnarray}}
\newcommand{\nn}{\nonumber}

\begin{document}
\title[]{A simple inert model  solves the little hierarchy problem\\
 and provides a dark matter candidate
}
\date{\today}
\author{F.\ Bazzocchi$^{\dag\ddag}$}
\author{M.\ Fabbrichesi$^{\ddag}$}
\affiliation{$^{\ddag}$INFN, Sezione di Trieste}
\affiliation{$^{\dag}$SISSA, via Bonomea 265, 34136 Trieste, Italy}

\begin{abstract}
\noindent   We discuss  a minimal extension to the standard model in which  two singlet scalar states that only interacts with the Higgs boson is added. Their masses and interaction strengths are fixed by the two requirements of canceling the one-loop quadratic corrections to the Higgs boson mass and providing a viable dark matter candidate. Direct detection of the lightest of these new states  in  nuclear scattering experiments is possible with a cross section within reach of  future experiments.
\end{abstract}

\pacs{11.30.Qc, 12.60Fr, 14.80.Bn, 95.35.+d}
\maketitle

Physics at LEP taught us how electroweak (EW) precision measurements prefer  a Higgs boson mass between 100 and 200 GeV~\cite{Higgsbounds} and---not having seen new particles beyond those of the standard model (SM)---a  cutoff $\Lambda$ for higher order operators encoding new physics larger than 5 TeV~\cite{Barbieri:2000gf}. Both these results seem to  be confirmed by the early runs of the LHC. In particular, the possible determination of the Higgs boson mass around 125 GeV~\cite{Higgsmass} comes in support to the first, while the absence of new states propagating below the  TeV scale~\cite{NPbounds} would  support  the second one. 

When taken together, the two statements above rise the problem of the little hierarchy: for the Higgs boson mass (and the electroweak vacuum expectation value)  to be in the 100 GeV range---that is, roughly between one and two orders of magnitude  smaller than the cutoff---quadratic renormalization effects must be canceled to an unnaturally high accuracy. This cancellation may either come from a symmetry---as it is the case in supersymmetric models---or be an accident in which the various terms conspire to cancel against each other. In the latter case,  the cancellation is best thought as the effect of a  dynamical mechanism, at work beyond the cutoff, which  arises from new physics that we do not know and  simulate by  fixing by hand some of the terms in the effective lagrangian. 

In this letter, we come back to the little hierarchy problem by following such an empirical approach and discuss a possible solution based on the presence of  inert scalar states~\cite{inerts}, that is, scalar particles only interacting with the Higgs boson (and gravity) which acquire no vacuum expectation value. 

The simplest realization is based on  the addition to the SM of  just two states: two real  scalars  $S_a$  and $S_b$ transforming as the  singlet representation of the EW gauge  group $SU(2)\times U(1)$ (and similarly not charged under the color group).

In addition we impose a $Z_2$ symmetry under which $S_{a,b}$ are odd and all the SM fields are even. In this way,  the new states couple  to the SM Higgs doublet only through quartic interactions in the scalar potential. 

By construction, our  model is inert and therefore we only look for solutions with $\vev{S_{a,b}}=0$, thus $Z_2$ is unbroken and after EW symmetry breaking the lightest singlet state can potentially be a viable  cold dark matter (DM) candidate. 

We  solve the little hierarchy problem by assuming that the Veltman condition~\cite{veltman} is satisfied, namely that the new scalar sector couples to the SM Higgs boson just so as to make  the one-loop quadratic divergences to the SM Higgs boson bare mass vanish (see \cite{noi} for an earlier application of the same idea in a non-inert model and \cite{wudka} for a multi-scalar implementation studying solutions different from ours). Once one-loop quadratic divergent terms are canceled, the Higgs boson mass is only  renormalized by one-loop finite  terms and  higher-loop corrections and can therefore be naturally smaller than the cutoff.

 In this approach, we may also ask whether there are natural solutions  for which the   bare masses of $S_{a,b}$  are of the same order of their one-loop quadratic correction, or, in other words, whether there are  solutions for which  $S_{a,b}$ has  a mass  of the order $m\gtrsim \Lambda/4 \pi$---that is, $m\gtrsim 700 $ GeV for $\Lambda \sim 10$ TeV.  Such  mass values would  then be natural thus removing the need of adding additional particles (fermions, in this case) to cancel {\it \`a la} Veltman the one-loop  quadratic divergences to the masses of $S_{a,b}$.  As we shall see, this is indeed the case.

The  lagrangian of the model  is given by the kinetic and Yukawa terms of the SM with a scalar potential given by
\bea
\label{pot}
V(H,S_a,S_b)&=&\mu_H^2 (H^\dag H)+ \mu^2_a S_a^2+\mu^2_b S_b^2\nn\\
& +&\lambda_1 (H^\dag H)^2 +
\lambda_{2_a} S_a^4+\lambda_{2_b} S_b^4\nn\\
&+& \lambda_{3_c} (H^\dag H ) S_a S_b \nn\\
&+& \lambda_{3_a} (H^\dag H ) S_a^2+ \lambda_{3_b} (H^\dag H ) S_b^2\nn\\
&+& \lambda_{4_a} S_a^3 S_b+  \lambda_{4_b} S_b^3 S_a+\lambda_{4_c} S_a^2 S_b^2 \,.
\eea
Notice that in \eq{pot} the quadratic term $S_a S_b$ is not present because   eliminated by a field redefinition.
A subset of the independent parameters of the potential in \eq{pot} is constrained  by requiring:
\begin{itemize}
\item[-] the  Veltman condition for the SM Higgs doublet renormalization;\\[-1.5em]
\item[-]   tree-level unitarity for $S_iS_j$ scattering;\\[-1.5em]
\item[-] correct relic density for the lightest $S_i$ as a cold DM candidate.
\end{itemize}
We impose no non-triviality condition.

The Veltman condition consists in the vanishing of the one-loop quadratic divergence contribution to the Higgs mass. The most unambiguous way to compute it is by using dimensional regularization (DR) and extract the pole in $D=2$. Other renormalization schemes, and in particular those with a sharp cutoff, give rise to scheme-dependent divergent  logarithmic terms which obscure the result. Accordingly we find
\be
\label{velt}
\delta \mu_H^2 \propto \frac{1}{16 \pi^2}\Big[  \frac{9}{4} g^2+\frac{3}{4}{g'}^2 +6 \lambda_1+\lambda_{3_a}+\lambda_{3_b}-12 y_t^2\Big]\,,
\ee
where $g$ and $g'$ are the EW gauge couplings and $y_t$ is the top Yukawa couplings---having neglected all the other much smaller Yukawa couplings.  
As usual the minimization of the scalar potential is obtained by imposing  that at the minimum $\vev{H}=v_W/\sqrt{2}$ and, as already mentioned,   by requiring  that $S_{a,b}$ be inert with vanishing vacuum expectation values. In this way,  we obtain three simple relations between the  masses of the  physical scalars, namely $h$,$S_a$ and $S_b$, and the parameters of the scalar potential $V(H,S_a,S_b)$:
\be
\lambda_1= \frac{m_h^2}{2 v_W^2}\,, \quad  \lambda_{3_{a}}+ \lambda_{3_{b}}= \frac{m_{S_{a}}^2+m_{S_{b}}^2- 2 \mu^2_{a}- 2 \mu^2_{b}}{v_W^2} \label{rel}
\ee
Accordingly, the Higgs boson mass quadratic divergent contribution becomes proportional to 
\bea
\label{velt2}
\delta \mu_H^2& \propto& \frac{1}{16 \pi^2 v_W^2} \Big[3 m_h^2 + 3 m_Z^2+ 6 m_W^2 +m_{S_a}^2+m_{S_b}^2 \Big.\nn \\
&-& \Big. 2 \mu^2_{a}-2 \mu^2_{b}-12 m_t^2\Big] \,.
\eea
Consistently we compute the one-loop  finite contributions to the Higgs boson mass using dimensional regularization with renormalization scale $\mu$.
The SM particle contributions are negligible while the new scalars, the mass of which is not fixed at this level, contribute with
\bea
\label{finite}
\delta m^2_h(\mu^2) &\simeq& \frac{1}{16 \pi^2} \Big[\lambda_{3_a} m^2_{S_a}\log\frac{m^2_{S_a}}{\mu^2} \Big. \nn \\  &+ & \Big. \lambda_{3_b} m^2_{S_b}\log\frac{m^2_{S_b}}{\mu^2}\Big]\, ,
\eea
where we are neglecting terms proportional to $\lambda_{3_c} v_W^2/(m^2_a - m^2_b)$ because they are small for the solutions we are interested in.

By  imposing the Veltman condition $\delta \mu_H^2=0$ we obtain the sum  $m_{S_a}^2+m_{S_b}^2$ in terms of the unknown  parameter $\mu_{a,b}^2$ and of the input physical quantities $m_h$, $m_t$, $m_W$ and $m_Z$ which, by substituting their experimental values (taking for $m_h$ the value of 125 GeV), yields
\be
m_{S_a}^2+m_{S_b}^2= 500^2+ 2 \mu_{a}^2+ 2 \mu_{b}^2\,,
\ee 
with an overall uncertainty---given by the neglected lighter quark masses contribution---of about 5\%.

Since $S_{a,b}$ are inert, $\mu_{a,b}^2 >0$ and we automatically have  a lower bound for $m_{S_a}^2+m_{S_b}^2$ given by $500^2$ GeV$^2$. Moreover, the   Higgs boson discovery put a constraint on the Higgs decay into light singlet scalars $h\to S_i S_i$,  which  gives an individual  lower bound for $m_{S_{a,b}}$ of  roughly 60 GeV---well below the mass range we are interested in.  

In addition,  the sum $\lambda_{3_a}+\lambda_{3_b}$ is  fixed by  \eq{rel} to be
\be
\lambda_{3_a}+\lambda_{3_b} = \frac{500^2}{v_W^2} \simeq 4.1\,, \label{4.1}
\ee
a rather large value which is however still within the perturbative regime  since $(\lambda_{3_{a,b}})^2/4\pi^2 < 1$ and the triviality bounds on the Higgs potential.

 Tree-level unitarity for the scattering of the additional scalars can be verified by examining the partial wave unitarity for the  two-particle amplitudes for energies $s\geq M_{W} ^2, M_Z^2$ .  We  write the  $J = 0$ partial wave amplitude $a_0$ in terms of the tree-level amplitude $T$ as
\be
a_0(s) = \frac{1}{32\pi}\int_{-1}^1 d\cos\theta\, T(s) =   \frac{1}{16\pi} F(\lambda_i)\,. \label{f}
\ee
We   can compute the amplitudes by using only the scalar potential because of  the equivalence theorem~\cite{Unitarity}. Accordingly, in the  high-energies regime, the only relevant contributions come from the quartic couplings in the  potential and $F$ in \eq{f} is  a function of the $\lambda_i$  in the potential in \eq{pot}. 

The combination of the  $\lambda_i$  entering in $F$ is constrained  by unitarity which requires $a_0(s)<1$. If we take $\lambda_1$, $\lambda_{3_a}$ and $\lambda_{3_b}$ according to \eq{rel} and \eq{4.1}, the unitarity constraint is satisfied by taking all the other $\lambda_i$ of order one.

The EW precision measurements are automatically satisfied by the singlets inert model because of the value of the Higgs boson mass and the non-interaction of the new states with the rest of the SM particles. 


The inert scalars $S_i$, being  gauge singlets, only interacts with the SM particles through  the Higgs boson $h$.  The point-like interaction $\lambda_{3_k}/2\, S_iS_j h h$  and the scattering mediated by $h$---both in the $s$ and  $t$ channels---contribute  to the cross section $S_iS_j\to hh$. The Higgs boson $h$ also  mediates  the scattering processes $S_iS_j\to f\bar{f}$, $S_iS_j\to W^+ W^-$, $S_iS_j\to Z Z$. 

\begin{figure}[t!]
\begin{center}
\includegraphics[width=3in]{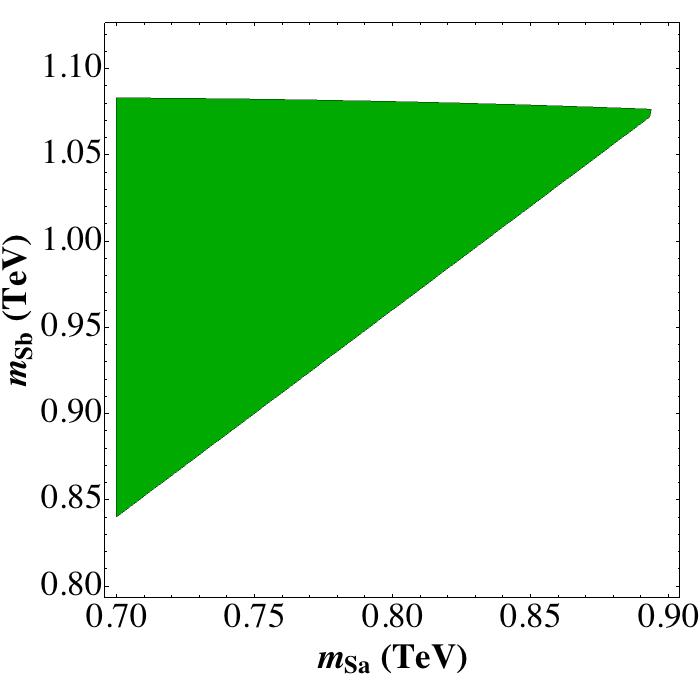}
\caption{\small   The allowed region for the singlet masses $m_{S_a}$ and $m_{S_b}$ satisfying all the constraints  discussed in the text. The upper line comes by requiring  finite one-loop corrections to the Higgs boson mass of order $100$ GeV when the lightest singlet state provides the correct amount of relic density in the universe. The diagonal line  corresponds to the condition that $S_b$ is sufficiently heavier than $S_a$  so as to allow neglecting coannihilation processes when computing the relic density. Finally the vertical line on the left is the lowest mass value for  ${S_a}$ which is still  natural  for an effective theory below $10$ TeV. 
\label{fig0}}
\end{center}
\end{figure}

It has been shown~\cite{Yaguna:2008hd} that a single inert singlet  that couples with the Higgs boson with a small coupling    is a realistic cold DM candidate with  a mass $\lesssim v_W$. In our case,  the lightest singlet  may account for the correct relic density in the opposite regime where  its mass is $ \gg v_W$ and its coupling with the Higgs boson   relatively large.  In this case,  fixing $S_a$ to be the lightest and assuming that  $S_b$  is sufficiently heavier not to take into account coannihiliation processes,  the  scattering amplitude is dominated by the pointlike  $S_aS_a\to hh$ vertex  which gives a contribution to the total cross section equal to
\be
\label{sigma}
\langle \sigma v \rangle  \simeq \frac{1}{16 \pi} \frac{\lambda_{3_a}^2}{m_{S_a}^2}\,.
\ee
To estimate the viability of $S_a$ as DM candidate,  we make use of the approximated analytical  solution~\cite{Srednicki:1988ce}. The relic abundance $n_{\mathrm{DM}}$  is written as 
\be
\label{exp}
\frac{n_{\mathrm{DM}}}{s}= \sqrt{\frac{180}{\pi g_{*}}}\frac{1}{M_{pl} T_f \langle \sigma  v\rangle}\,,
\ee
where $M_{pl}$ is the Planck mass, $T_f$ is the freeze-out  temperature, which for our and similar  candidates  is given by $m_{S_a}/T_f\sim 26$. The constant $g_*= 106.75+2$  counts the number of  SM degrees of freedom in thermal equilibrium plus the additional degrees of freedom related to the singlets, $s$ is their total entropy density. Current data fit within the standard cosmological model  give a relic abundance with $\Omega_{\mathrm{DM}} h^2=0.112\pm 0.006$~\cite{PDG} which corresponds to 
\be
\label{value}
\frac{n_{\mathrm{DM}}}{s}= \frac{(0.40\pm 0.02)}{10^9 \, m_{S_a}/\mbox{GeV}}\,. \label{exp2}
\ee
Under the hypothesis  that the two singlets have a mass $\geq 800$ GeV not to be protected at the one-loop level,  that  the $S_b$ decays mainly through $S_b\to S_a h h$, and that the finite contribution in \eq{finite} is of order the Higgs boson mass, \eqs{exp}{exp2}    determine an allowed  region in the  singlet masses space as shown in Fig.~\ref{fig0} and roughly delimited by
$$
0.7 \leq  m_{S_a}\;  \mbox{(TeV)} \leq 0.9 \quad \mbox{and} \quad
0.8\leq  m_{S_b}\;  \mbox{(TeV)}  \leq 1.1 \,. 
$$
The uncertainty here is dominated by the approximations built in the analytical expression given by \eq{exp} which can be estimated to be of the order of $10\%$.

Notice that  had we introduced only one singlet state, the solution of the little hierarchy problem would have given it a mass and a coupling to the Higgs boson leading to a relic abundance too large to agree with current determinations. This is the motivation behind the introduction of two rather than just one additional scalar.

We verified---by using the software of~\cite{PPPC}---that photon and charged particle fluxes are two or more  orders of magnitude smaller than those corresponding to a typical EW scale DM candidate. For this reason, there seems to be little hope of testing the model in such indirect DM searches. 

\begin{figure}[t!]
\begin{center}
\includegraphics[width=3.4in]{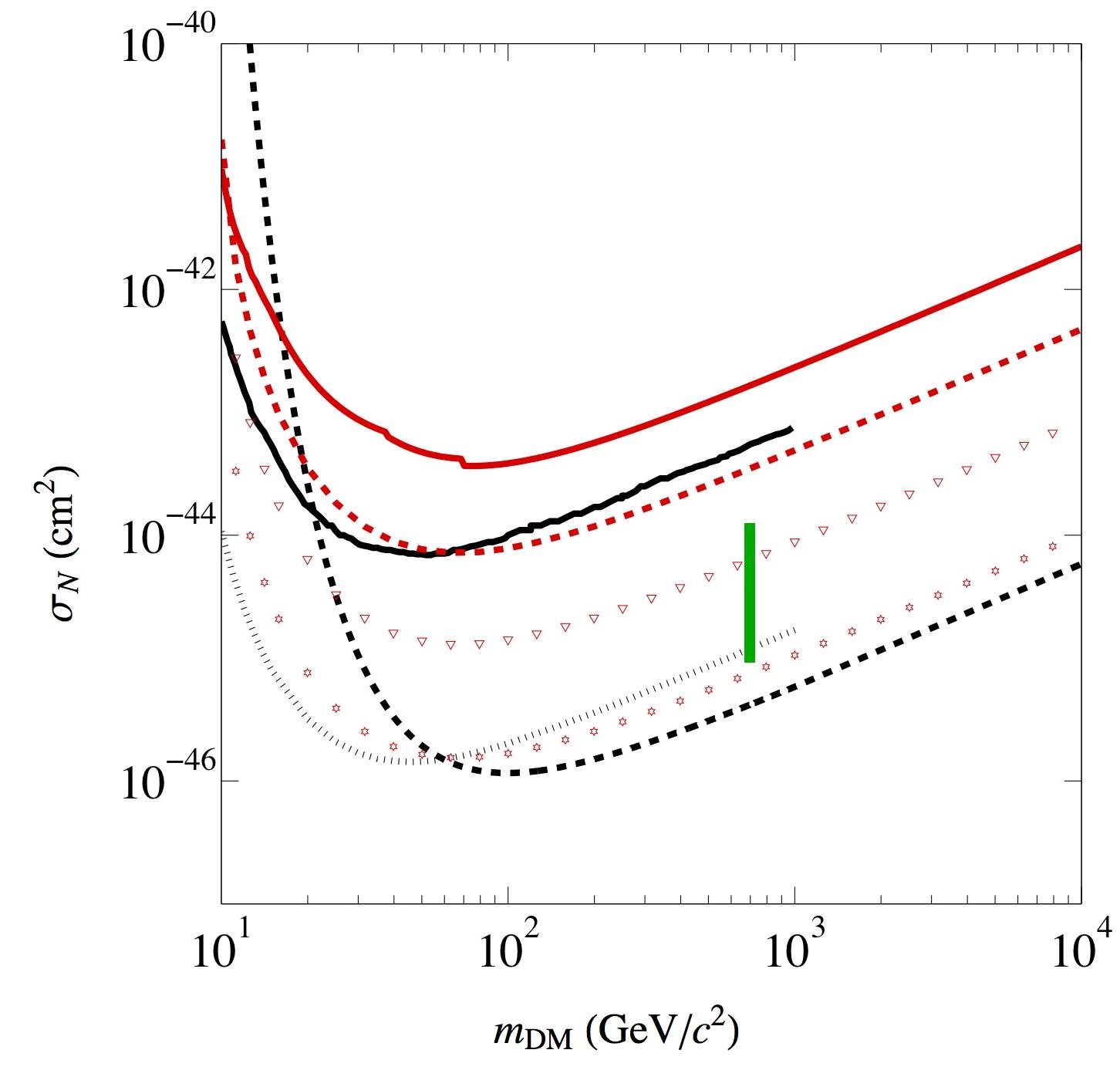}
\caption{\small   Spin independent cross section per nucleon versus  DM candidate masses~\cite{DMT}. The black (red) solid line corresponds to the  XENON100 (CDMSII) data. Black points and the black dashed  line are   the projections for upgraded XENON100  and XENON1T, respectively.  The red dashed line, down triangles and stars correspond to different projections for SCDMS. The green vertical line  is the prediction of the inert  inert model discussed in this letter.
\label{fig}}
\end{center}
\end{figure}

 On the other hand,  there exists
a  possibility of detecting the inert scalar $S_a$  in nuclear scattering experiments. The $\lambda_3$ quartic term in \eq{pot} gives rise also to the three fields interaction $SSh$ which yields the effective singlet-nucleon vertex
\be
f_N  \frac{\lambda_3 m_N}{m_h^2} S_a S_a \,\bar{\psi}_N \psi_N\,.
\ee
The (non-relativistic) cross section for the process  is given by~\cite{sigma}
\be
\sigma_N = f_N^2 m_N^2 \frac{\lambda_3^2}{4 \pi} \left( \frac{m_r}{m_{S_a} m^2_h} \right)^2 \,,
\ee
where $m_r$ is the reduced mass for the system which is, to a vary good approximation  in our case, equal to the nucleon mass $m_N$; the  factor $f_N$ contains many uncertainties due to the computation of the nuclear matrix elements and it can vary  from  0.3 to 0.6~\cite{nucleon}. Substituting the values we have found for our model, we obtain, depending on the choice of parameters within the given uncertainties, a cross section 
$\sigma_N$ between $ 10^{-45}$  and  $10^{-44}\mbox{cm}^2$, a value within reach of the next generation of experiments  (see Fig.~\ref{fig}).

Let us conclude by  commenting on other possible inert models obtained by using larger $SU(2)$ representations. The  inert doublet  model~\cite{inerts} has  been studied in the past    because it  provided a solution to the  little hierarchy problem in the regime of a large Higgs boson mass, a scenario  now ruled out by latest  LHC data~\cite{Higgsmass}. 

More recently,  inert models  with doublet or higher representations have   been studied because they provide viable DM candidates~\cite{inertoni}. 
Once we require that an inert model satisfies all the constraints discussed in this letter, namely Veltman condition for the Higgs boson mass, correct relic density for the lightest neutral state, finite contributions to the Higgs boson mass of order $100$ GeV, the case of only one inert state transforming as a non-trivial complex representation   of $SU(2)$, let us call it $R$, resembles the case of one singlet and does not allow for any solution. 

The reason is that  the Veltman condition in the case of a non-trvial representation  modifies the couplings in the potential only by a multiplicity factor, thus implying that the  quartic coupling $hh RR$  remains too large unless  very big ($d_R \sim 10$) representations are invoked. Then, by requiring the correct relic density, this turns in a mass value   larger than 1 TeV and, as a consequence, the finite contributions to the Higgs boson mass are too large.  Moreover, the mass splittings among the components of $R$  have to be controlled so as not to affect EW precision measurements, implying that not all the   coannihilation effects   can be neglected. 

In conclusion, even by promoting the inert states to be in a non-trivial complex representation of $SU(2)$, we still need at least two of them, as in the case of the model described in this letter, and no advantage seems to be gained.

\acknowledgments

 We thank P.\ Ullio and N.\ Fornengo for explaining to us some aspects of  DM physics. MF thanks SISSA for the hospitality.


\end{document}